\documentclass[aps,pra,10pt,twocolumn,notitlepage]{revtex4-2} 
\usepackage{amsfonts}
\usepackage{amsmath}
\usepackage{graphicx,xcolor}
\usepackage{hyperref}

\begin{document}


\title{Testing eigenstate thermalization hypothesis on small scale quantum processors}
\author{Maanav Srihari} 
\affiliation{School of Physics, Indian Institute of Science Education and Research, Thiruvananthapuram, Kerala, India 695551}
\affiliation{Center for High Performance Computing, Indian Institute of Science Education and Research, Thiruvananthapuram, Kerala, India 695551}

\author{Anil Shaji}
\affiliation{School of Physics, Indian Institute of Science Education and Research, Thiruvananthapuram, Kerala, India 695551}
\affiliation{Center for High Performance Computing, Indian Institute of Science Education and Research, Thiruvananthapuram, Kerala, India 695551}

\begin{abstract}
The Eigenstate Thermalization Hypothesis (ETH) is a framework for discussing thermal behavior originating from chaotic dynamics in isolated many-body quantum systems. The PXP model, where certain states do not thermalize, has been compared with the Sachdev-Ye Kitaev (SYK) model, which is believed to be fully thermalizing. A gate-based quantum circuit approach is utilized to simulate time evolution and compute the Out-of-Time-Ordered Correlator (OTOC), a measure of the extent of chaos. Considering restrictions on implementing SYK on gate-based hardware, a simplified model called Spin-XY4 (SXY4), which has a thermal behavior similar to SYK, is tested. An alternate method, which optimizes control on an analog quantum device with the GRAPE (GRadient Ascent Pulse Engineering) algorithm, is also utilized to simulate the SYK model.
\end{abstract}

\maketitle

\section{Introduction \label{intro}}

Substantial progress has been made over the last century or more in understanding the dynamics of macroscopic systems, starting from the laws of physics that apply at microscopic scale
. However, there are still several open questions to be resolved, one of which is understanding the process of thermalization, particularly when the system under consideration is not too large. Thermalization refers to the evolution of a system to a state that is not dependent on initial conditions with expectation values of relevant observables also being independent of the initial conditions. Two important concepts used to explain thermalization are dynamical chaos and ergodicity. Chaos refers to the idea that the phase space trajectories of a complex system are extremely sensitive to the initial conditions, and even an infinitesimal change can cause the trajectories to diverge radically \cite{strogatz2018nonlinear}. Ergodicity refers to the ability of a system to visit all possible points in its configuration space \cite{moore2015ergodic, klein1952ergodic}. 

When the underlying microscopic dynamics is quantum mechanical, one has to address the additional problem that the definition and characterization of chaos does not directly translate to quantum systems. For instance, the position-momentum uncertainty relation does not allow the definition of a phase space or trajectories in them. For ergodic classical systems, time averages can be found by computing averages for an ensemble. In an isolated quantum mechanical system, though, time averages depend on the initial conditions because the evolution is deterministic and is governed by the Schrodinger equation \cite{nandkishore2015many, deutsch2018eigenstate}. Expectation values of observables, therefore, have to be computed with respect to the single time evolved state and not with respect to an ensemble. 

The Eigenstate Thermalization Hypothesis (ETH) \cite{srednicki1994chaos, deutsch1991quantum} is one approach to explaining how quantum mechanical expectation values of various observables computed with respect to a single time-evolved state can be equal to the ensemble averages for equivalent classical systems. ETH is built In close analogy with an ergodic classical system for which the point in phase space corresponding to its instantaneous state explores all parts of the available phase space in time. For an equivalent quantum system, the ETH stipulates that steady states of the Hamiltonian are superpositions of all available states such that the probabilities obtained from the corresponding amplitudes of the superposition furnish the appropriate ensemble. In other words, energy eigenstates of quantum chaotic systems themselves work like ensembles \cite{deutsch2018eigenstate, rigol2008thermalization}. This makes the notion of thermalization inherent to the system and is a much stronger statement because instead of the phase point eventually exploring all of the available phase space, individual steady states have support on the entire set of available states at any given time. 

In this paper, we do a comparative study of two interesting quantum many-body Hamiltonians, the PXP model \cite{bernien2017probing, ebadi2021quantum, lesanovsky2012interacting} and the Sachdev-Ye-Kitaev (SYK) model \cite{sachdev1993gapless, RevModPhys.94.035004} by simulating them on quantum processors. Both models are known to be chaotic yet possess distinct properties. It is known that PXP model exhibits the presence of non-thermal eigenstates in the form of ``many-body'' scars that cause long time oscillations in the probability \cite{ebadi2021quantum, bernien2017probing, serbyn2021quantum, turner2018weak}, similar to scars of classical periodic orbits in the wavefunction of a quantum particle inside a stadium billiard \cite{heller1984bound}. The scar can prevent the thermalization of the eigenstates with probability amplitudes accumulating around a few states. The SYK model is an example of a Hamiltonian said to be strongly chaotic, with all its eigenstates expected to be thermal in nature \cite{maldacena2016bound}. We also study the Spin-XY4 (SXY4) model \cite{hanada2024model}, which is similar in structure to the SYK model and is believed to have thermalization characteristics.

The size of the Hamiltonians of quantum many-body systems increases exponentially with the number of particles that describe that system. In addition, strong interactions can restrict simplifying assumptions that may make the dynamics generated by the Hamiltonian efficiently computable on classical computers. Typically, classical numerical simulations of such systems are therefore possible only for relatively small system sizes. In contrast, quantum computers offer the potential for scalable simulation of the dynamics, where the number of qubits required or the computational depth does not increase exponentially with system size and qubit-qubit interactions can be engineered to mimic the interactions governing the many-body system under consideration. Quantum computers are broadly classified as gate-based or analog. Gate-based devices can implement simulations by breaking down the unitary governing the evolution of the many-body system into predefined sets of gates that can be chained together sequentially in a quantum circuit. Analog devices implement simulations by directly affecting the controls on the Hamiltonian that governs the processor.

In this paper, we implement gate-based simulations of the PXP \cite{desaules2024robust},  SYK \cite{asaduzzaman2024sachdev, luo2019quantum} and SXY4 models. We also perform analog simulation of the SYK model. In simulations of the three models signatures of thermalization of eigenstates, as well as that of scars, are shown. The gate-based approach is additionally used to compute the Out-of-Time-Ordered Correlator (OTOC) for the SYK model \cite{asaduzzaman2024sachdev, vermersch2019probing}, which is a measure of the extent of thermalization in many-body quantum systems \cite{swingle2018unscrambling, xu2024scrambling}. We restrict our simulations to the dynamics generated only by the interaction terms of these Hamiltonians by going to a suitable interaction picture and assuming other conditions under which the one-body terms in the Hamiltonian can be neglected. This allows us to compare the eigenstates and their thermalization with the micro-canonical ensemble for corresponding classical systems. 

Outline of this paper is as follows: in Section \ref{eth} we review ETH and Section \ref{pxp} details the simulation of the PXP model on a gate-based approach. Note that analog simulation of the PXP model showing the scar state has been done previously \cite{wurtz2023aquila}. Section \ref{syk} details the simulation of the SYK model on both a gate-based and analog processor-based approach while Section \ref{sxy} is on the simulation of the SXY4 model. Section \ref{otoc} contains the computation of the OTOC values using a gate-based simulation. Finally, our observations and conclusions are in Section \ref{discus}.

\section{Eigenstate Thermalization Hypothesis \label{eth}}

Consider an isolated many-body quantum system starting from an arbitrary state $|\Psi\rangle = \sum_i c_i |E_i\rangle$, written in the basis of energy eigenstates $|E_i\rangle$, with $E_i$ denoting the energy. The long-time average of an observable $A$ with respect to the state $|\Psi (t) \rangle$ that is evolving according to the  Schrodinger equation is given by \cite{rigol2008thermalization, deutsch2018eigenstate}, 
\[\langle A(t)\rangle = \sum_i |c_i|^2 \langle E_i|A|E_i\rangle,\]
We see that the long-time average depends on the initial state of the system through the probabilities, $|c_i|^2$. If the system undergoes thermalization, it should settle to an average that corresponds to an appropriate ensemble, irrespective of the initial state. For the case of an isolated quantum system, the appropriate one is the microcanonical ensemble. The average of $A$ with respect to the microcanonical ensemble is given by \cite{rigol2008thermalization},
\[\langle A\rangle_{\rm mc}(E_0) = \frac{1}{\mathcal{N}_{E_0, \Delta}}\sum_i \langle E_i|A|E_i \rangle.\]
Here, the average is over all allowed states with energy $E_i$ that lie inside the shell of half-width $\Delta$ surrounding the mean energy of the initial state given by $E_0$. $\mathcal{N}_{E_0, \Delta}$ is the number of states that lie inside the shell \cite{rigol2008thermalization}.

The two methods of computing the average need reconciliation because the long-time average from Schrodinger evolution depends explicitly on the initial state, while the microcanonical average does not. Such a resolution is possible if one assumes that the expectation value of the observable $A$ in an energy eigenstate $|E_i\rangle$, denoted by $A_{ii} \equiv \langle E_i| A | E_i\rangle$ is identical to the microcanonical average at a mean energy $E_i$ \cite{rigol2008thermalization, deutsch2018eigenstate},
\[A_{ii} = \langle A\rangle_{mc} (E_i).\]
This means that the state $|E_i\rangle$ implicitly acts as the ensemble. Time evolution then assumes a secondary role in manifesting thermal behavior, unlike in a classical system where thermalization occurs due to the exploration of its configuration space \cite{rigol2008thermalization}. In the quantum case, time evolution has to drive the energy eigenstate to a state whose probability amplitudes furnish the appropriate ensemble as stipulated by ETH. 

 Quantum mechanical chaos can be characterized on the basis of chaotic behavior in the classical limit. For quantum systems whose classical limit is chaotic, it was found that the distribution of energy level separations follows the same statistics as random matrices. This is the Bohigas-Giannoni-Schmit (BGS) conjecture \cite{bohigas1984characterization}. The ETH was originally found to explicitly hold true for such systems in \cite{srednicki1994chaos, deutsch1991quantum}. Similarly, in the context of many-body Hamiltonians, the SYK model is predicted to be strongly chaotic, and therefore, the eigenstates are expected to thermalize \cite{maldacena2016bound}. In certain systems, not all energy eigenstates may exhibit thermalization. This flexibility allows this class of systems to exhibit a weak violation of ETH, where the presence of eigenstates that do not follow thermal behavior leads to interesting dynamics and final states far removed from thermal states. An example of this was discovered experimentally in an array of Rydberg atoms, where the contribution of non-thermal eigenstates to certain states of the Rydberg atom system leads to a revival of the probability of that state \cite{bernien2017probing, ebadi2021quantum}, rather than relaxing to thermal states. Such violations of ETH are particularly interesting in the context of quantum materials, especially when the ground state shows non-thermal behavior either due to scarring or any other reason. One can then expect potentially exotic macroscopic properties arising from the non-thermal ground state. 

 \begin{figure*}[!htb]
  \centering
  \resizebox{17.5cm}{5.1 cm}{\includegraphics{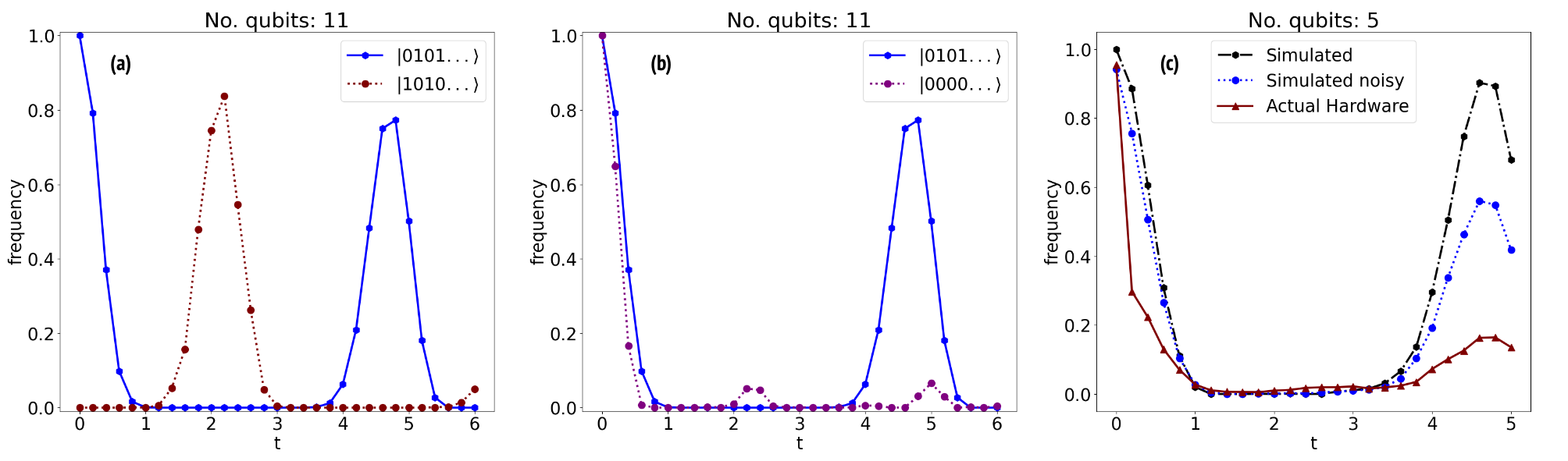}}
  \caption{(a) The oscillation between $|\mathbb{Z}_2\rangle$ states, when starting from $|0101 \ldots \rangle$ is seen in a computation run on an 11 qubit quantum simulator with no noise.   (b) Time evolution of the initial states $|000 \ldots \rangle$ and $|\mathbb{Z}_2\rangle = |0101 \ldots \rangle$ are compared in the 11 qubit quantum simulator without noise. While the probability for the $|Z_2\rangle$ state shows a recurrence after a finite interval of time, the probability decays to a very low value $\sim 2^{-11}$ in the case of the initial $|000\ldots\rangle$ state.  (c) Comparison of probability revivals for the 5-qubit $\mathbb{Z}_2$ state $|01010\rangle$ between computations run on the 5-qubit quantum simulator with and without noise and the same computation done on the IBM Brisbane system is shown. The quantum hardware does perform quite well, showing the recurrence of the state after a finite interval of time.}
  \label{fig:PXP}
\end{figure*}

In the following, we study quantum many-body systems in which each individual subsystem is a qubit with $|0\rangle $ and $|1\rangle$ denoting the computational basis states of each qubit. The SYK, PXP, and SXY4 Hamiltonians that we study apply to these multi-qubit systems. As mentioned earlier, we assume that we work in a suitable interaction picture with model parameters chosen such that the respective Hamiltonians contain only multi-qubit interaction terms \cite{wurtz2023aquila,lesanovsky2012interacting,maldacena2016remarks}. Since all external drive terms have been transformed away, the multi-qubit systems we simulate can be effectively treated as isolated systems. Moreover, since single qubit terms are missing from the Hamiltonians, both $|0\rangle$ and $|1\rangle$ states are degenerate in energy for each qubit. This also means that all possible computational basis states for an $N$-qubit system denoted as $\{ |\vec{x}\rangle \}_0^{2^N}$ are degenerate and form the set of allowed basis states for the system. This implies that if ETH holds, then any initial state with a fixed energy that is assumed to be the same as the energies of each of the basis states should evolve into an equal superposition of all the computational basis states, furnished through the respective probability amplitudes that are all equal in magnitude to each other, like the microcanonical ensemble. It may be noted that if the Hamiltonians that are considered have any additional symmetry like parity etc., the superposition is confined to only those computational basis states with the same symmetry as the initial state.

\section{The PXP model \label{pxp}}

The PXP model is defined on a one-dimensional array of qubits and is described by the Hamiltonian, 
\begin{equation}
  H_{\rm PXP} = \sum_i P_{i-1} \sigma^x_i P_{i+1}
\end{equation}
Here $P_i = (\openone_2 - \sigma^z_i)/2$ are the projection operators to the ground state of the $i^{\rm th}$ qubit, and $\sigma^x_i$s denotes the Pauli $X$ operator acting on the $i^{\rm th}$ qubit. The model has been implemented directly on the analog quantum processor QuEra Aquila \cite{wurtz2023aquila} that uses Rydberg atoms as qubits. Scarring is observed in this system when starting with the $|\mathbb{Z}_2\rangle$ states, given by $|1010...\rangle$ or $|0101...\rangle$. When these states are time evolved, they switch between each other instead of thermalizing. Other initial states, however, do show thermalizing behavior with signatures of the $|Z_2\rangle$ states appearing as residual scars in cases where the initial state has an overlap with one or both $|Z_2\rangle$ states~\cite{serbyn2021quantum, turner2018weak, lesanovsky2012interacting}. If an initial state like $|000...\rangle$ that lacks overlap with the $|Z_2\rangle$ states is chosen, then the time evolution is thermalizing, leading to states with uniform probability on all available basis states with little or no revival. 

This PXP Hamiltonian can also be simulated on a gate-based quantum computer. We chose to do such a simulation of the dynamics generated by the PXP Hamiltonian using IBM Qiskit~\cite{qiskit2024, desaules2024robust}. The first step in the simulation is replacing the $P_i$ operator in $H_{\rm PXP}$ with a string of Pauli operators leading to, 
\begin{equation}
  \label{eq:2}
  P_{i-1} \sigma^x_i P_{i+1} = \frac{1}{4}(\sigma^x_i - \sigma^z_{i-1} \sigma^x_i - \sigma^x_i \sigma^z_{i+1} + \sigma^z_{i-1} \sigma^z_{i+1})
\end{equation}
We used an open boundary condition, and the boundary terms are $\sigma^x_0 P_1$ and $P_{N-1} \sigma^x_N$, with $N$ being the number of qubits. To construct the terms of the Hamiltonian according to equation \eqref{eq:2}, \verb|SparsePauliOp| from Qiskit is utilized. The terms generated are then arranged into collections containing operators commuting with each other using the function \verb|group_commuting|. Once the Hamiltonian is encoded onto the gate-based quantum processor in this manner, the corresponding unitary time evolution operator for a short interval of time is generated using \verb|PauliEvolutionGate|. This time evolution operator is concatenated several times to obtain a Trotter approximation for the overall time evolution operator. If a Hamiltonian is written as $H = \sum_k H_k$, with $H_k$ consisting of Pauli strings, the Trotter approximation with $M$ steps leads to the following approximation for the unitary time evolution~\cite{nielsen2010quantum, desaules2024robust}:
\[e^{-iHt} \approx \bigg[\prod_k e^{-iH_k t / M} \bigg]^M,\]
with higher values of $M$ leading to better accuracies. 

For an 11-qubit circuit, starting from the $\mathbb{Z}_2$ state, the oscillation associated with scarring was observed using the \verb|StateVectorSampler| which is a classical simulator from Qiskit. These oscillations are shown in Fig~\ref{fig:PXP} (a). However, starting with $|000...\rangle$ does not lead to such oscillations, and the probabilities for obtaining each of the $2^{11} = 2048$ states becomes uniform as shown in  Fig~\ref{fig:PXP} (b) where the evolution of one of the two $|Z_2\rangle$ states is also shown for comparison. We repeated the same simulation on a 5-qubit circuit running on the quantum hardware \verb|ibm_brisbane| \cite{qiskit2024} and in this case also, the oscillatory behavior is seen for the $|Z_2\rangle$. A comparison of runs using $|Z_2\rangle$ initial state on quantum simulators with and without errors as well on actual quantum hardware is  shown in Fig~\ref{fig:PXP} (c). We see that the results from the quantum hardware is promising, as far as scaling up to more qubits in future quantum processors goes, since the revival of the state is clearly seen.

\begin{figure}[h!]
  \centering
  \resizebox{7 cm}{14 cm}{\includegraphics{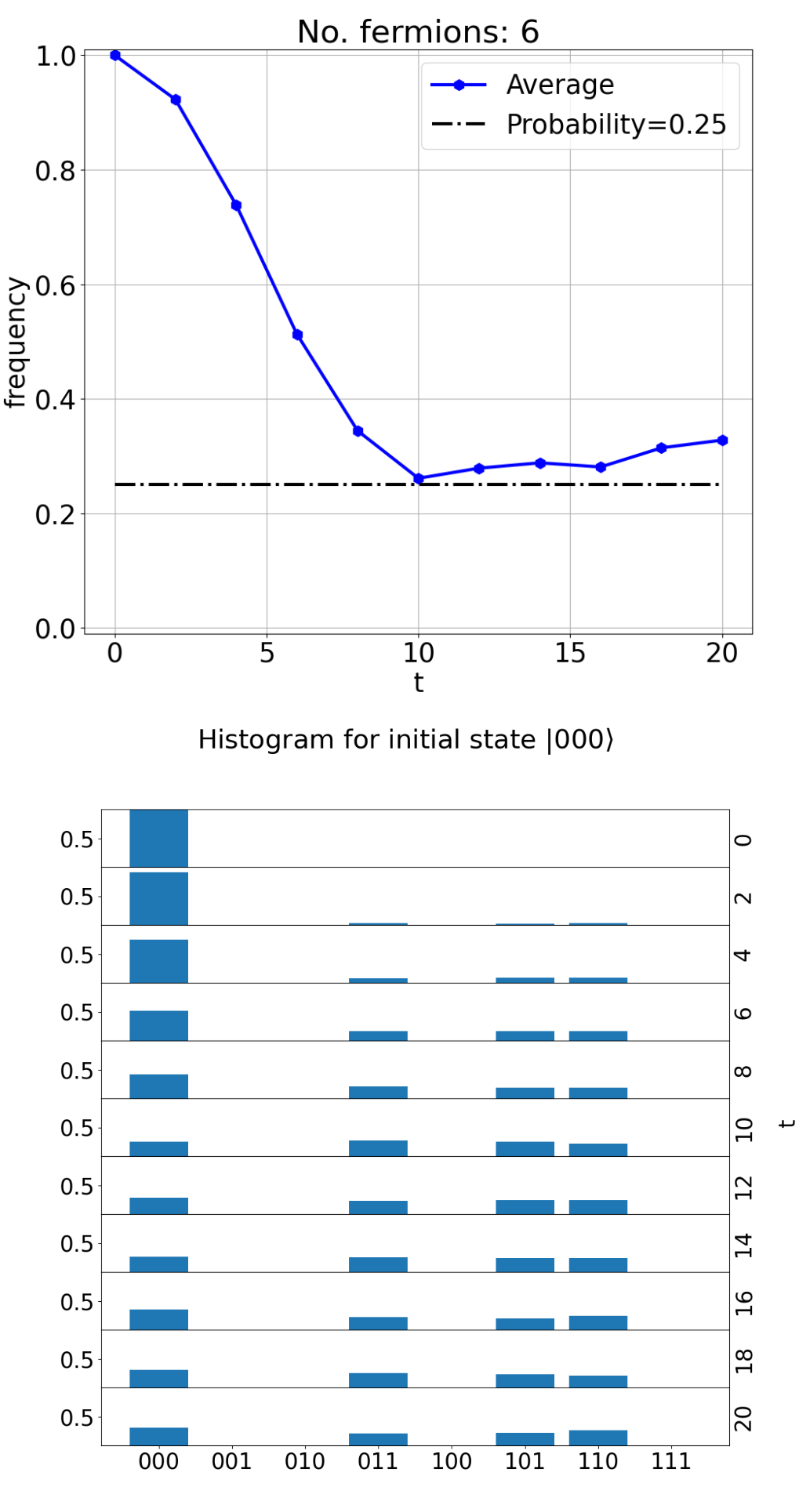}}
  \caption{The figure on top shows the average frequency which corresponds to the survival probability of the initial state, taken over multiple sets of $J_{ijkl}$ and $\eta_{ijkl}$ values for SYK evolution, starting from $|000\rangle$. The probability saturates around $\sim 1/2^2 = 0.25$, which is shown by the black dotted line. The figure at the bottom shows the corresponding histograms for the probabilities of each of the eight computational basis states at each SYK time-step $\tau$. In the stack of histograms, time flows downward, and we see that the state evolves to one having equal probabilities on all basis states having an even number of $1$'s. This is because the SYK Hamiltonian preserves parity.}
  \label{fig:SYK1}
\end{figure}

\section{The SYK model \label{syk}}

The SYK Hamiltonian is of interest because it is known to be one among the class of models which are the fastest scramblers of quantum information, thereby likely to be fully thermalizing according to ETH \cite{maldacena2016bound}. It was initially proposed by Sachdev and Ye to describe the strange-metal phase in high-temperature superconductors \cite{sachdev1993gapless, RevModPhys.94.035004}. Later, it was simplified to model the properties of black holes in the context of scrambling dynamics \cite{maldacena2016remarks}. The Hamiltonian of the model without any single particle terms is,
\begin{equation}
  H_{\rm SYK} = \sum_{i,j,k,l} J_{ijkl} \chi_i \chi_j \chi_k \chi_l
\end{equation}
This Hamiltonian contains a four-body interaction between Majorana fermions given by the operators $\chi_i$ and the coefficients, $J_{ijkl}$, are random numbers chosen from a normal distribution. One can see that the interactions are quite complex, with randomly distributed strengths that contribute to the capacity of the Hamiltonian to generate chaotic dynamics. We simulate the evolution of randomly chosen initial states using both gate-based and analog quantum computers to verify that irrespective of the initial state and for all random choices of coefficients, the states do thermalize. To make the simulations tractable on the available quantum hardware we simplify the model by modifying the Hamiltonian to 
\[H'_{\rm SYK} = \sum_{i,j,k,l} \eta_{ijkl} J_{ijkl} \chi_i \chi_j \chi_k \chi_l,\] 
where the parameter $\eta_{ijkl}$ is set to 0 with a probability $p$ and is set to the value 1 with a probability $1-p$. This allows us to make the Hamiltonian sparse enough to enable the simulation by choosing $p$ close to unity \cite{tezuka2023binary, garcia2021sparse}.

\subsection{Gate-based simulation}

The Majorana fermion operators appearing in the SYK Hamiltonian need to be converted to a form that can be implemented using qubits while preserving the anti-commutation relation $\{\chi_i,\chi_j\} = \chi_i\chi_j + \chi_j\chi_i = \delta_{ij}$. This mapping to qubits is done via the Jordan-Wigner transformation, using strings of Pauli operators as \cite{asaduzzaman2024sachdev}:
\begin{equation}
  \label{eq:3}
  \chi_{2k-1} = \bigg(\prod_{i}^{k-1} \sigma^z_i\bigg) \sigma^x_k, \qquad
  \chi_{2k} = \bigg(\prod_{i}^{k-1} \sigma^z_i\bigg) \sigma^y_k 
\end{equation}
From equation \eqref{eq:3}, it is seen that one needs $N$ qubits to simulate $2N$ Majorana fermions. Making the Hamiltonian sparse reduces the complexity further to allow for the simulation of larger systems. 

\begin{figure}[h!]
  \centering
  \includegraphics[width=6.5cm]{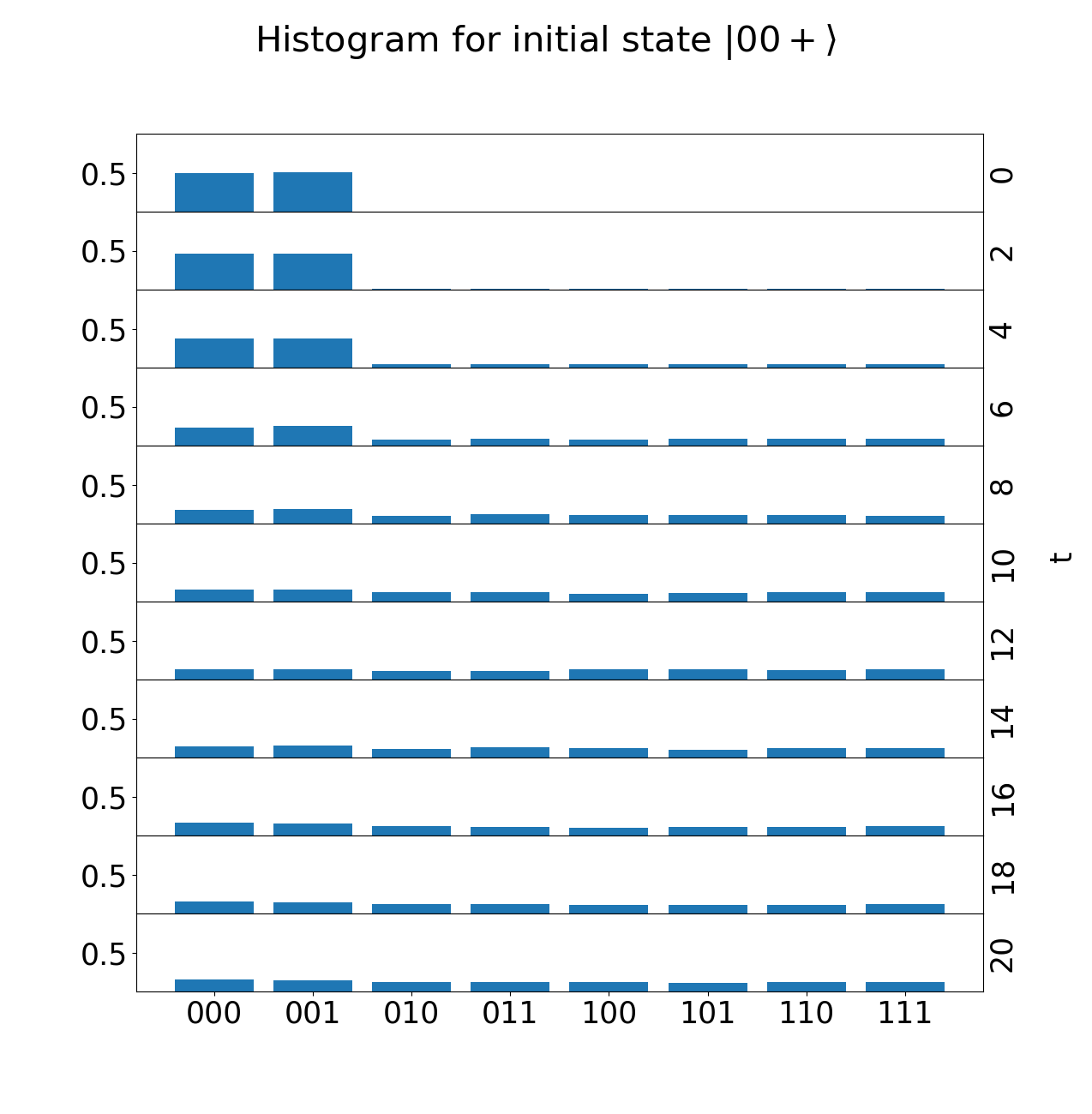}
  \caption{The occupation probability (frequency) of each of the eight computational basis states at each time step $\tau$ when parity in the $\sigma^z$ basis is left unconstrained by starting with $|00+\rangle$ state for the SYK evolution run on a quantum simulator. The frequencies shown are averaged over multiple sets of $J_{ijkl}$ and $\eta_{ijkl}$ values. The eigenstate evolves to an equal superposition of all eight basis states with the probability amplitude of each saturating to  $\sim 1/2^3 = 0.125$, indicating thermalization.}
  \label{fig:SYK2}
\end{figure}

As in the case of the implementation of the PXP model, from the Qiskit class \verb|SparsePauliOp| and its function \verb|group_commuting|, along with \verb|PauliEvolutionGate|, were used to create the quantum circuit for the gate-based simulation which also uses the Trotter approximation \cite{asaduzzaman2024sachdev, luo2019quantum}. This model is tested on a local classical simulator with \verb|StateVectorSampler|. Evolution of the 3-qubit (6 fermions) initial state $|000\rangle$ under the SYK Hamiltonian is plotted in figure \ref{fig:SYK1}. We see that the probability of the system remaining in the initial state decays to $1/2^2 = 0.25$ indicating thermalization. Also shown in Fig.~\ref{fig:SYK1} is a histogram that shows the time evolution of the probabilities associated with each of the eight possible computational basis states starting from the $|000\rangle$ state at time $t=0$ which is shown in the top panel with time increasing as we go down the stack of histograms in the figure. In both the panels of Fig.~\ref{fig:SYK1}, we have averaged over several trials with randomly chosen values of $J_{ijkl}$ and $\eta_{ijkl}$. We note that only four of the eight possible basis states are populated by the time evolution, and thermalization with equal probability seems to happen within the sub-space defined by these four states. However, within this sub-space, there is no tell-tale sign of a scar which would be a higher probability for one or more of these states.

 The SYK model conserves total spin parity with respect to the operator $\hat{Z} = \prod_i \sigma^z_i$ in the computational basis, so starting from a state having even (odd) number of qubits in the $|1\rangle$ the time evolution would preserve the number of qubits in state $|1\rangle$ \cite{bhore2023deep}. For the 3-qubit case (6 fermions) starting with $|000\rangle$, the possible states that it can evolve to include $|000\rangle$, $|011\rangle$, $|101\rangle$ and $|110\rangle$, which together form a four-dimensional subspace.  For the averaged evolution, the probability of finding the system in each state is $0.25$. When the fixed parity of the initial state is removed by starting from $|00+\rangle$, $|+\rangle = (1/2)(|0\rangle + |1\rangle)$, the evolved state shows equal support on all eight computational basis states with the probability for each going to $1/2^3 = 0.125$ as shown in figure \ref{fig:SYK2}. 
 
\begin{figure}[!htb]
  \centering
  \includegraphics[width=6.0 cm]{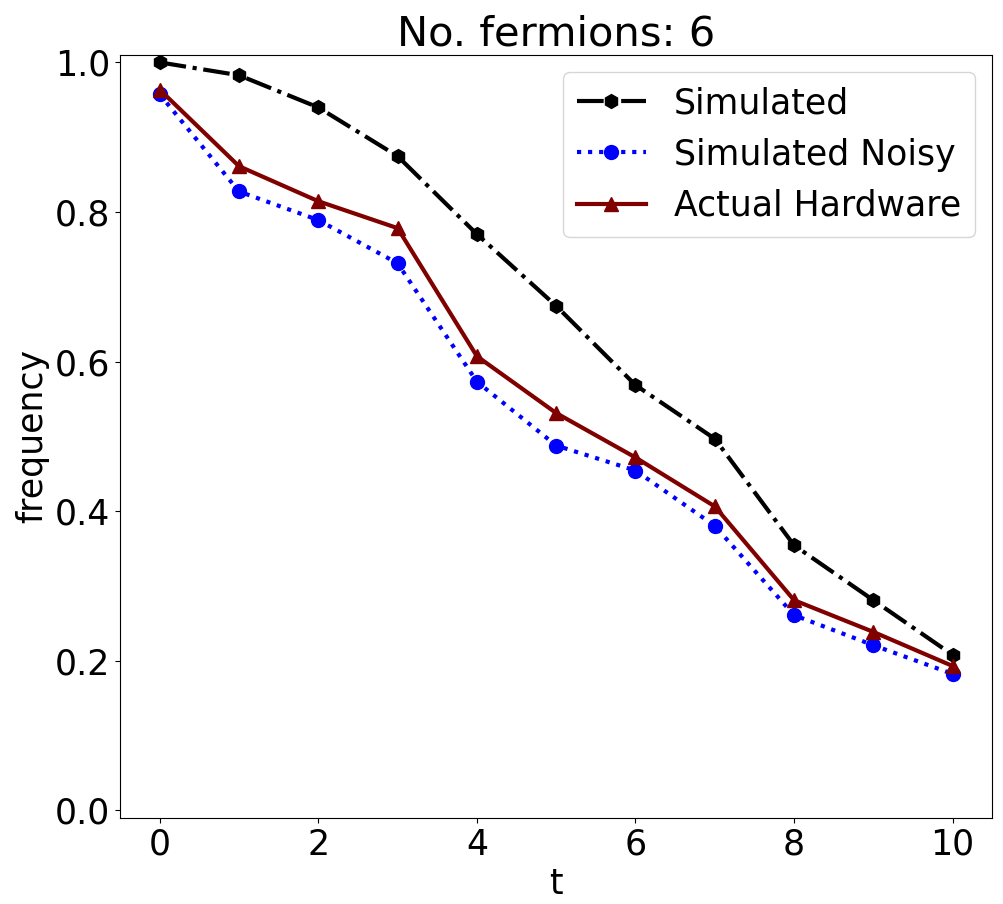}
  \caption{Comparison of the survival probability of the initial state $|000\rangle$ under SYK evolution when computed using quantum simulators with and without noise as well as on quantum hardware. We see good agreement between all three showing the potential for scaling up when quantum computers with more qubits become available. Scaling up of the classical simulations is however limited by computational complexity.  \label{SYKHarware}}
\end{figure}

 We see that on the quantum simulator the SKY model shows behavior that is faithful to the ETH. We further performed the same computations both on quantum simulators with noise added as well as on quantum hardware, namely \verb|ibm_brisbane|. The results of these computational runs are shown in Fig.~\ref{SYKHarware}, and we see that close agreement with the expected behavior is obtained showing the potential for scaling up when quantum computers with more qubits become available. Note that there is no averaging over different choices of $J_{ijkl}$ and $\eta_{ijkl}$ in the simulations run on quantum hardware. 

\subsection{Analog simulation}

While the PXP Hamiltonian is naturally implemented on QuEra Aquila, which is a Rydberg atom-based quantum processor that can be accessed through the Amazon Braket cloud platform \cite{wurtz2023aquila}, implementation of the SYK Hamiltonian on the same platform is not so straightforward. In QuEra Aquila, the excited states of the neutral atom qubits are Rydberg levels. The Rydberg-atom qubits can be precisely arranged with a user-defined configuration on a lattice using optical tweezers. They can then be excited by a global laser drive that produces Rabi oscillations. The amplitude, phase, and detuning of the drive can be controlled by the user. Qubit-qubit interactions are implemented through van der Waals' forces. The Hamiltonian that describes the system is,
\begin{eqnarray}
  H_{\rm Ryd} & = &  \frac{\Omega(t)}{2} \sum_i (e^{i\phi(t)} |0\rangle_i\langle 1| + h.c.)   - \Delta(t)\sum_i |1\rangle_i\langle 1| \nonumber \\
 && \qquad  + \sum_{i < j} \frac{C_6}{|r_i - r_j|^6}
  |1\rangle_i\langle 1| \otimes |1\rangle_j\langle 1|
\end{eqnarray}
Here $\Omega(t)$ is the drive amplitude, $\phi(t)$ the phase, and $\Delta(t)$ the detuning. $|0\rangle_i$ ($|1\rangle_i$) denotes the ground (excited) state of the qubit at position $i$ \cite{wurtz2023aquila, bernien2017probing, ebadi2021quantum}.

Evolution under the SYK Hamiltonian is broken down into time steps $\tau$, with a unitary matrix corresponding to each; $U_{SYK}(\tau) = exp(-iH_{SYK}\tau)$. The method devised here involves engineering a simulation of these unitary operators on the Rydberg atom processor, starting from a fixed initial state. The drive amplitude $\Omega(t)$ is fixed at a constant value $2\pi \times 0.75 MHz$ ($2\pi \times 2.5 MHz$ is the maximum value allowed on QuEra Aquila \cite{wurtz2023aquila}). The phase $\phi(t)$ is set to $0$. Since only global driving is possible as of now, asymmetry is introduced by positioning the qubits on the lattice such that spacing between them is not identical while taking care that no qubit is in the Rydberg blockade radius of another. Runtime that is possible on the processor (and its simulator) is fixed to $4\mu s$. Implementation of each  $U_{SYK}(\tau)$ is then found by optimizing the requisite detuning, $\Delta(t)$.

Optimization is done using the GRAPE algorithm \cite{khaneja2005optimal} using the package \verb|QuantumControl.jl| \cite{goerz2022quantum}, available in the Julia programming language. It uses the method of semi-automatic differentiation via the package \verb|Zygote.jl| \cite{Zygote.jl-2018} to perform the computations for the optimization procedure. The process starts with an  initial guess function for $\Delta(t)$, discretized as a piecewise constant pulse. The figure of merit to be minimized in the case of quantum control is the fidelity error, given by $\epsilon = 1 - |\langle \psi_{target}|\psi_{final}\rangle|^2$, with $|\psi_{target}\rangle$ being the target state that should be obtained if $U_{SYK}(\tau)$ acts on the initial state while $|\psi_{final}\rangle$ is the actual state obtained after applying the initial guess for the control pulse. The guess function is varied piece by piece to find the function that minimizes $\epsilon$. 

\begin{figure}[!htb]
  \centering
  \includegraphics[width=7cm]{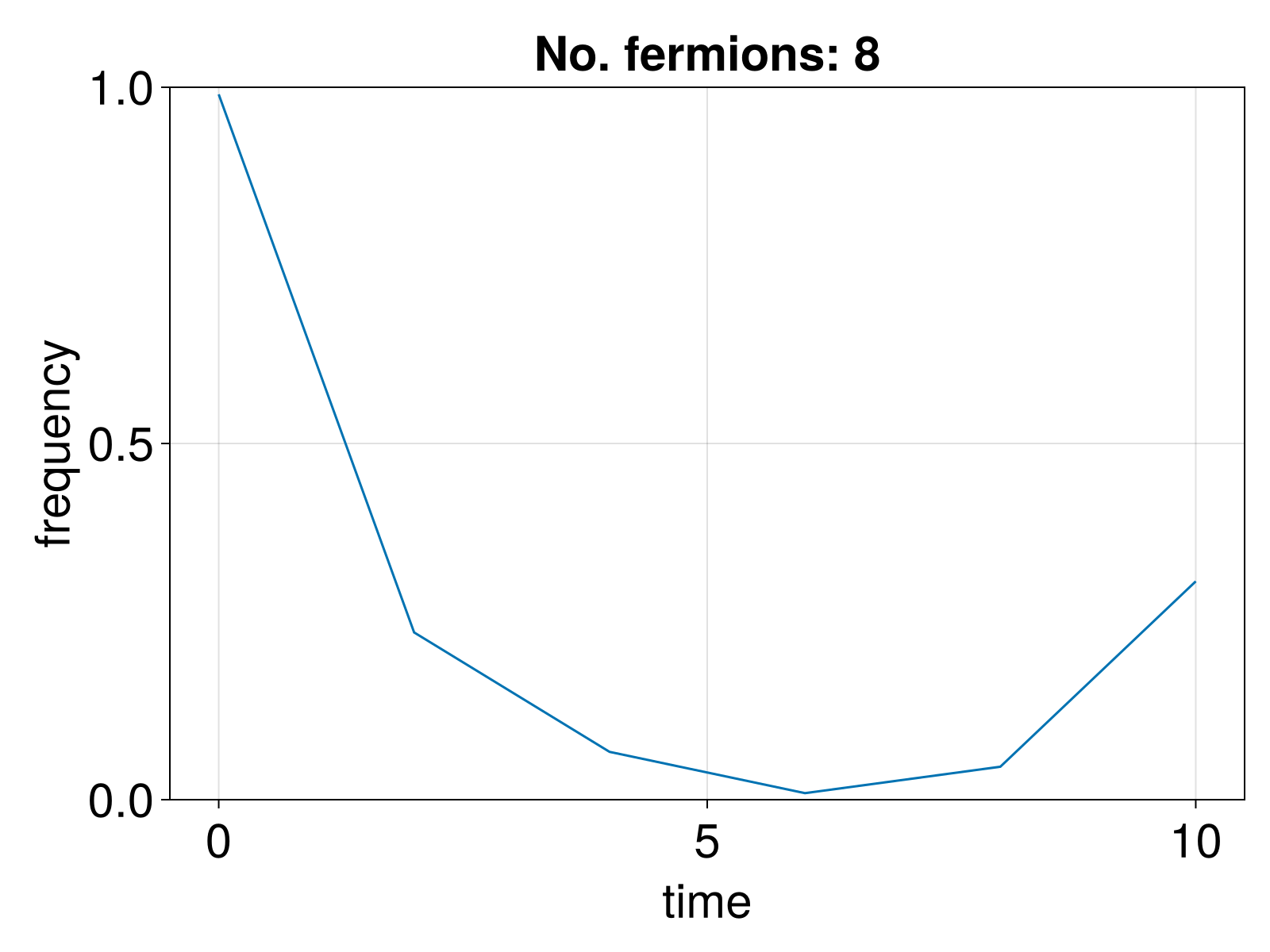}
  \includegraphics[width=8cm]{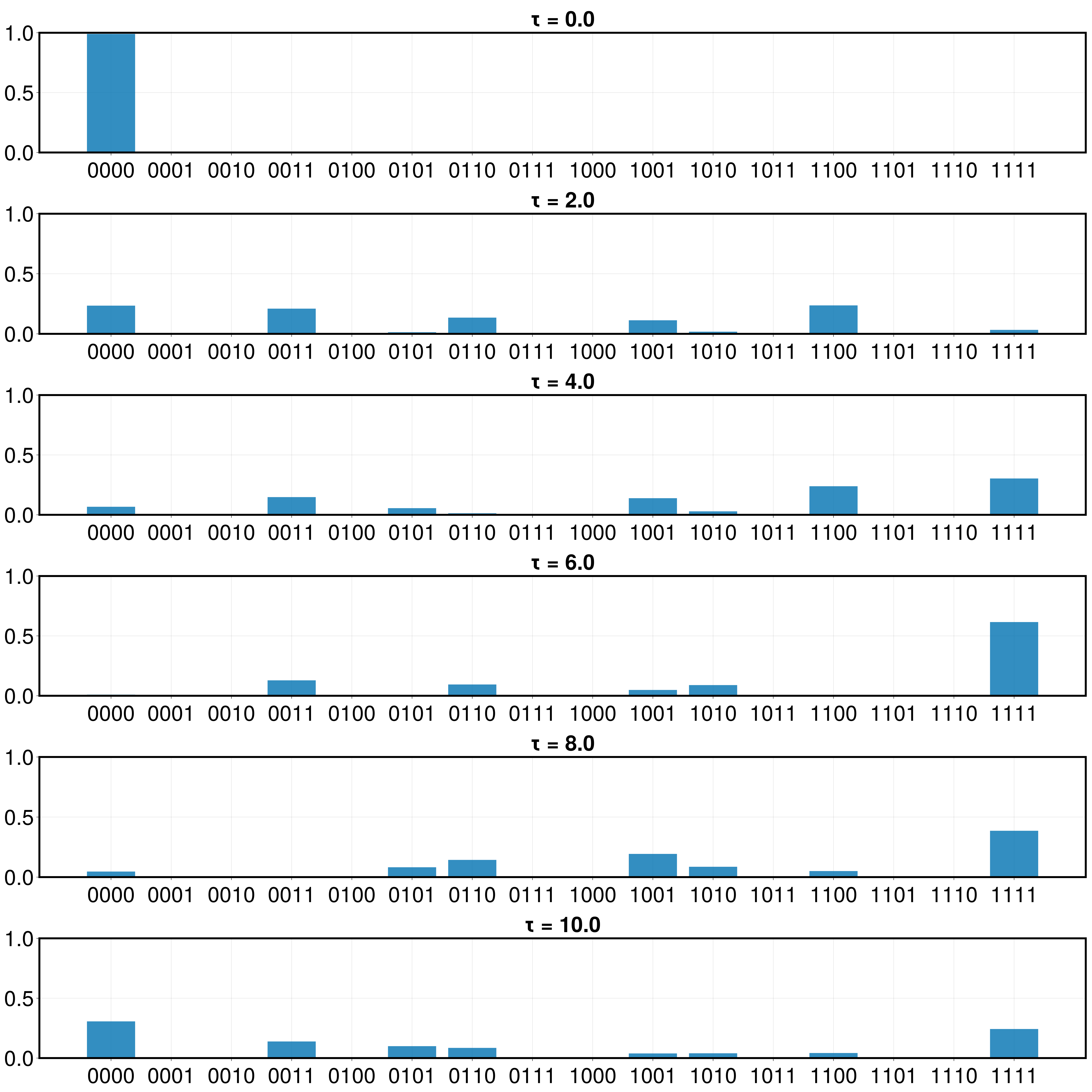}
  \caption{Time evolution of the $|0000\rangle$ state under the SYK Hamiltonian computed on the quantum simulator available from QuEra is shown in the figure on top. The one at the bottom shows the  corresponding histogram of probabilities of each qubit state at different SYK time-steps, $\tau$. \label{syk1}}
\end{figure}

Results of a run of the SYK model on a classical simulation of the QuEra device with 4 qubits (8 fermions) are shown in Fig.~\ref{syk1}. We see qualitative agreement with the gate-based computation. The computational errors are greater than in the previous since SYK Hamiltonian is not a natural fit for the processor. However, the successful simulation of the PXP model indicates that custom-built analog simulators can be an effective way forward, at least in the near-term, for simulating solid state models on quantum processors at a scale that makes it considerably more effective than classical computations that are limited by the size of the system.  


\section{The SXY4 model \label{sxy}}

While being important for understanding the nature of thermalization, the SYK model is challenging to implement on the current generation of gate-based quantum computing devices, which are restricted by the limited range of qubit-qubit interactions and smaller circuit depths. With this in mind, it is instructive to look at the properties of an alternate model with similar chaos and thermalization properties as SYK, in the form of SXY4 \cite{hanada2024model}. This model is much simpler to implement on hardware because it does away with the anti-commutation relation of the Majorana fermion operators, and considers a new set of operators as,
\begin{equation}
    \label{eq:4}
    O_{2k-1} = \sigma^x_k, \qquad O_{2k} = \sigma^y_k
\end{equation}
The Hamiltonian of the model is,
\begin{equation}
  \label{eq:5}
  H_{\rm SXY4} = \sum_{i,j,k,l} (i)^{\phi_{ijkl}} J_{ijkl} O_i O_j O_k O_l.
\end{equation}
Only a maximum of four qubits at a time are required to simulate a single term of the Hamiltonian, unlike SYK, where all the qubits need to be considered at once in the worst-case scenario, involving interaction between the first qubit, indexed by $1$, and the last qubit, indexed by $N$ as can be seen from \eqref{eq:3}. In equation \eqref{eq:5}, the exponent $\phi_{ijkl}$ counts the number of times that in a given term of $H_{\rm SXY4}$ the $\sigma^x$ and $\sigma^y$ operators appear next to each other. For example, $\phi_{1357} = 0$, since all operators in the corresponding term of $H_{\rm SXY4}$ are $\sigma^x$. On the other hand, $\phi_{1235} = 1$, and $\phi_{1234} = 2$. The term $i^{\phi_{ijkl}}$ is makes the Hamiltonian  Hermitian \cite{hanada2024model}. 

\begin{figure}[h!]
  \centering
  \includegraphics[width=7.6cm]{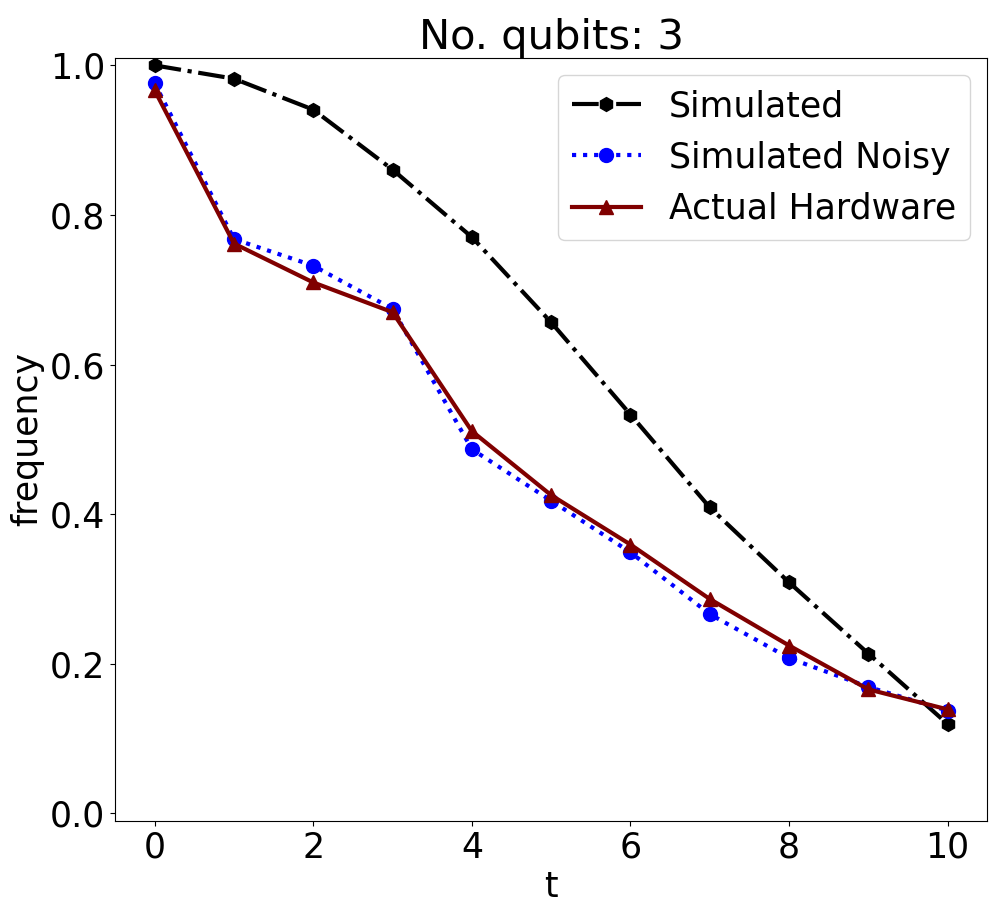}
  \caption{Evolution of the survival probability of the state $|000\rangle$ under SXY4 Hamiltonian. Computational runs on quantum simulators with and without noise models, as well as run on the IBM  Kyiv processor, are shown. We see good agreement between the simulated results and actual runs, and in both cases, indicate thermalization of the eigenstate. \label{sxy4-ibm}}
\end{figure}

We simulate the SXY4 Hamiltonian on a gate-based quantum processor. The eigenstates are seen to thermalize as with the SYK model. The SXY4 Hamiltonian also preserves parity \cite{bhore2023deep}, and starting with $|000\rangle$ state leads to a probability of $0.25$ for each of the four possible states. When the fixed parity is removed by starting from $|00+\rangle$, the probability changes to $0.125$ for each of the eight possible states. The computational runs on quantum simulators yield results almost identical to that for the SYK model shown in Figs.~\ref{fig:SYK1} and \ref{fig:SYK2}. In addition, this model is run on quantum hardware, \verb|ibm_kyiv| \cite{qiskit2024}. The results of the run for a fixed choice of $J_{ijkl}$ are shown in Fig.~\ref{sxy4-ibm}. 

\section{Out-of-Time-Ordered Correlator (OTOC) \label{otoc}}

We now look at a means of quantifying the degree of chaos that arises from the time evolution generated by the three model Hamiltonians that we have considered since the chaotic nature of the evolution in the classical limit is intimately connected with the thermalization of eigenstates. The typical measure of chaos in classical systems is the Lyapunov exponent that shows the extent of divergence of trajectories with initial conditions that are infinitesimally close to each other \cite{strogatz2018nonlinear}. An analog is found in the Out-of-Time-Ordered Correlator (OTOC) for quantum systems \cite{maldacena2016bound, xu2024scrambling, swingle2018unscrambling}. This quantity is based on the Out-of-Time-Ordered Commutator which is defined as
\begin{equation}
  \label{eq:6}
  C(t) = \langle[W(t), V(0)]^\dagger [W(t), V(0)]\rangle
\end{equation}
Where $W(t)$ and $V(0)$ are local operators corresponding to two different times. For a many-body quantum system, these operators act on sites that are well separated from each other, implying that $C(0) = 0$, as $W(0)$ and $V(0)$ commute. For the cases we consider, these operators are effectively single qubit operators at $t=0$. Due to interaction terms in the Hamiltonian, information initially localized at the site where $W$ acts spreads to the location where $V$ acts, and as a result, $C$ will no longer be equal to $0$. If one assumes further that the operators considered are Hermitian and unitary, equation \eqref{eq:6} is reduced to $C(t) = 2(1-Re[\langle W(t)V(0)W(t)V(0)\rangle])$. From here, a new quantity can be defined as
\begin{equation}
  O(t) = \langle W(t)V(0)W(t)V(0) \rangle
\end{equation}
which is the OTOC. Note that $O(t) = 1$ at $t=0$ since $C(0) = 0$, and $O(t)$ is expected to decay with time for a thermalizing system. Since $O(t)$ lacks time-ordering, measuring it directly in an experiment requires the ability to precisely reverse the time evolution back to the initial conditions, which is not possible in general. Alternatively, it can be measured using statistical correlations \cite{vermersch2019probing}. In a gate-based quantum processor, the measurement involves being operated on by a random unitary, followed by the time evolution generated by the Hamiltonian being considered and measurements. The exact method of computation of the OTOC is given below
\begin{equation}
  \label{eq:7}
  O(t) = \frac{1}{\overline{\langle W(t) \rangle^2}_{u,k_0}} \overline{\langle W(t) \rangle_{u,k_0}}\overline{\langle VW(t)V \rangle_{u,k_0}}.
\end{equation}
Here, $k_0$ denotes a starting state, and the overline indicates averaging over several random unitaries $u$. 

\begin{figure*}[!htb]
  \resizebox{14 cm}{11.5 cm}{\includegraphics{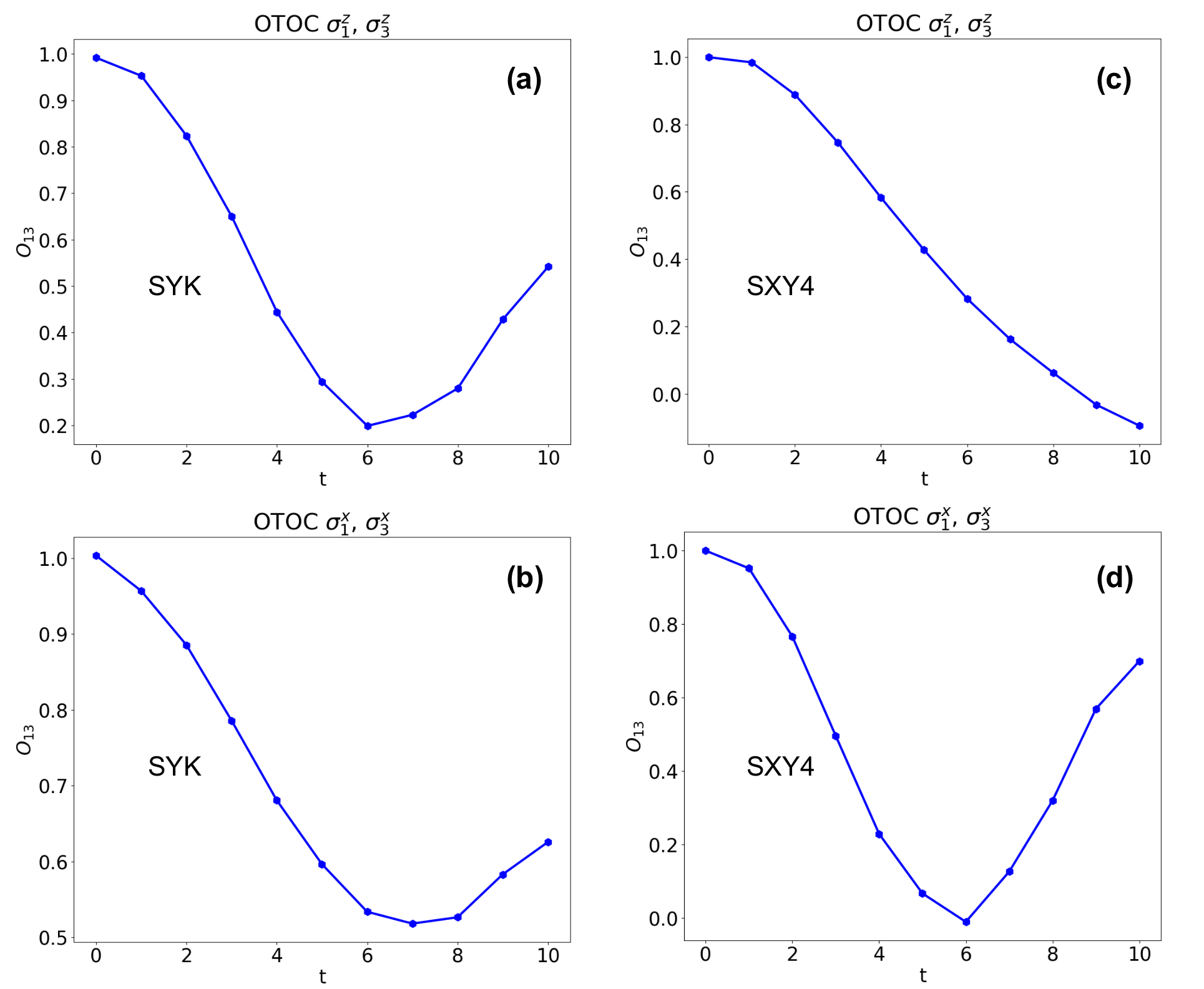}}
  \caption{ Figures (a) and (c) show the time dependence of the OTOC between $\sigma^z_1(t)$ at time $t$ acting on qubit-1 and $\sigma^z_3(0)$ at time $0$ acting on qubit-3 when the qubits are evolving under the SYK and SXY4 Hamiltonians respectively. Figures (b) and (d) show the OTOC between $\sigma^x_1 (t)$ on qubit-1 at time $t$ and $\sigma^x_3(0)$ on qubit-3 at time $0$ again for both models. We see that apart from recurrences due to the small number of qubits considered, the OTOCs show a strong tendency to decay with time.  The computations were performed on a gate-based quantum simulator. \label{fig:otoc}}
  \label{fig:OTOCSXY4}
\end{figure*}

Using equation \eqref{eq:7}, the OTOC was computed for different time steps for both the SYK and SXY4 models for a 3-qubit case on the quantum simulator using \verb|StateVectorSampler| in Qiskit. The local operator pairs considered were $\sigma^x_1$, $\sigma^x_3$ and $\sigma^z_1$, $\sigma^z_3$ (denoted as $O_{13}$ in the plots of Fig.~\ref{fig:otoc}. In both of these cases, the value decays with time, with recurrences seen at later times due to the small number of qubits considered. At initial times, we see that the OTOCs in all cases considered show a strong tendency to decrease, indicating that the evolution is strongly chaotic, leading to rapid thermalization of eigenstates. 

\section{Discussion \label{discus}}

The PXP and SYK models are known to be quantum chaotic with respect to the BGS conjecture \cite{serbyn2021quantum, maldacena2016bound}, with the statistics of energy level separation matching that of random matrices. In PXP evolution, when described using the terminology of classical trajectories, the $\mathbb{Z}_2$ states correspond to unstable periodic orbits in otherwise chaotic surroundings \cite{ho2019periodic}. The SYK model, on the other hand, is known to saturate the bound based on the OTOC \cite{maldacena2016bound} corresponding to completely chaotic dynamics. While both PXP and SYK Hamiltonians are believed to be chaotic based on different approaches, it is interesting to note the difference in their respective thermalization behaviors that are manifestations of the underlying chaos. The PXP model shows weak violation of the ETH because of the presence of scars in the form of $\mathbb{Z}_2$ oscillations. The SYK model does not appear to show any such properties when plotting the time evolution, and the same is confirmed by the decaying OTOC value, in the limited-size simulations performed here. Many-body scarring offers a way to create exotic many-body ground states where the system does not settle into a thermal state by virtue of oscillations in probability of the non-thermal eigenstates. We also studied the SXY4 model and showed that its behavior is very similar to that of SYK. In particular, with respect to the time evolution of the OTOC, the SYK and SXY4 shared striking similarities. 

In addition to exploring ETH and its violations due to quantum scars, another objective of our work was to implement simulations of the solid-state models in presently available quantum processors. We have shown that such simulations can be effectively done and show the potential to scale up. In addition to the anticipated windfall from being able to simulate solid-state models at scales much larger than what can be done using classical computation, we see that such quantum simulations can be used to explore more fundamental questions like thermalization. For instance, given the complexity of the SYK model, it is not obvious for all choices of the couplings and the coefficients $\eta_{ijkl}$ that control the sparsity of the Hamiltonian, the model will satisfy ETH. In the limit where the Hamiltonian is extremely sparse, one expects ETH not to hold anymore for the SYK model. Alternatively, scars may also appear for specific choices of the couplings in a non-sparse case. Both these questions become computationally accessible when they can be simulated on quantum computers with sufficiently large enough number of qubits and good connectivity.  

Inspired by the successful simulation of the PXP model on the Rydberg atom-based QuEra analog quantum processor, we showed that such analog simulations could be extended to other models like SYK also. This approach presents a promising alternative to the simulation of unitary dynamics on gate-based systems. Here, one is directly engineering the Hamiltonian on the analog processor, and it offers a more direct way to implement such simulations. We were able to achieve this using suitable, well-established algorithms to design the controls. Note that algorithms like GRAPE may not scale favorably if one has to find the required pulses for implementing such Hamiltonians on a larger number of qubits. However, more powerful techniques like machine learning and AI-based methods can be pressed into service to address such issues. Such approaches, which are planned as future work, can make the analog simulation approach scale more efficiently and effectively compared to gate-based approaches.

\acknowledgments 
M.~S. and A.~S. thank MeITY QCAL cohort 2 for grant of credits on AWS enabling the use of the BraKet service.  A.~S.~was supported in part by QuEST grant No Q-113 of the Department of Science and Technology, Government of India.  The authors acknowledge the centre for high performance computing of IISER TVM for the use of the HPC cluster, {\em Padmanabha}. 

\bibliography{Refs}

\end{document}